# A Multivariate Hierarchical Bayesian Framework for Healthcare Predictions with Application to Medical Home Study in The Department of Veteran Affairs

Issac Shams, Saeede Ajorlou, and Kai Yang

*Abstract—* Recently the patient-centered medical home (PCMH) model has become a popular approach to deliver better care to patients. Current research shows that the most important key for succession of this method is to make balance between healthcare supply and demand. Without such balance in clinical supply and demand, issues such as excessive under and over utilization of physicians, long waiting time for receiving the appropriate treatment, and non-continuity of care will eliminate many advantages of the medical home strategy. To reach this end we need to have information about both supply and demand in healthcare system. Healthcare supply can be calculated easily based on head counts and available hours which is offered by professionals for a specific time period while healthcare demand is not easy to calculate, and it is affected by some healthcare, diagnostic and demographic attributes. In this paper, by extending the hierarchical generalized linear model to include multivariate responses, we develop a clinical workload prediction model for care portfolio demands in a Bayesian framework. Our analyses of a recent data from Veteran Health Administration indicate that our prediction model works for clinical data with high performance.

## I. INTRODUCTION

Healthcare system is a complex multi level system in which the primary care is the first point of contact between patients and the systems. Typically primary care providers deliver the majority of care that patients receive in their lifetime. If a patient's health care needs cannot be satisfied at primary care level, the patient is referred to an appropriate specialty care unit. For the whole healthcare system, timely access to care, continuity of care, comprehensiveness, or concern for the entire patient rather than one organ system, and coordination among all parts of the system are four pillars of care, Starfield [1], Forrest and Starfield [2]. Rust et. al. [3] report that the inability to get a timely appointment to a primary care physician increases the likelihood of patients visiting the emergency department. This hinders the appropriate management of chronic diseases that could have been effectively treated in a primary care setting. It is also important that patients see their own physicians in order to maintain continuity of care. Continuity of care is considered one of the hallmarks of primary care. Gill and Mainous [4] point to several studies which show that patients who regularly see their own physicians are more satisfied with their care, more likely to take medications correctly, more likely to have problems correctly identified by their physicians and less likely to be hospitalized. Gill et. al. [5] show a link between lack of continuity and increased emergency department use. Traditional form of primary care is featured by primary care physicians (PCP), in which each primary care physician has a designated set of patients, called a patient panel, Murray et. al. [6]. In this setting, if the patient panel is the right size, a PCP will be able to see patients when their needs arise, rather than referring them to another day or another PCP. Clearly this PCP-panel pair is the main vehicle, which ensures the continuity of care. The size and composition of a panel will determine the amount of healthcare workload (or healthcare demand) in provider hours or minutes within a given period (typically a year). Green at. al. [7] investigate the link between panel size and the probability of 'overflow' or extra work for a physician. They conclude that the supply of provider hours has to be sufficiently higher than demand to offset the effect of variability. There are many ways to determine a patient panel.

In recent years, patient centered medical home (PCMH) has become a popular model for providing health care services, especially at primary care level. Patient centered medical home is a team based service and each team consists of a group of healthcare professionals, such as physician, nurse practi- tioner, clerk, social worker, nutritionist, pharmacist and so on. Team members and the patient share the patient records so everyone sees the same records. The medical home concept originated during 1960's in pediatrics Carrier et. al. [8]. Presently, the PCMH model has been practiced by many hospitals and medical centers, Bitton et. al. [9], Friedberg et. al. [10] and its performance has been evaluated by many studies, Nutting et. al. [11], Jaen et. al. [12], and Crabtree et. al. [13]. A good patient panel design and management methodology is even more critical for PCMH model than the traditional PCP model for the following reasons:

- In the traditional single PCP model, the healthcare supply is the total available hours of physician time within a given period (typically a year) by a PCP, and the healthcare demand is the total requested

*Resrach supported by National Science Foundation.

Issac Shams is with the Department of Industrial and Systems Engineering, Wayne State University, Detroit, MI, 48202 USA. (e-mail: er7671@wayne.edu).

Saeede Ajorlou, is with the Department of Industrial and Systems Engineering, Wayne State University, Detroit, MI, 48202 USA. (e-mail: er7212@wayne.edu).

Kai Yang, is with the Department of Industrial and Systems Engineering, Wayne State University, Detroit, MI, 48202 USA. (e-mail: ac4505@wayne.edu).

physician hours generated by the patients in the panel. The healthcare supply can be treated as deterministic, and the healthcare demand as a random variable. In PCMH model, the healthcare supply is a portfolio of total available hours by various members in a team within a particular period, (e.g., total physician time, total nurse time, total clerk time, etc.), the healthcare demand is a portfolio of demand requested by the patients in the

- Patient panel to PCMH team members, the healthcare supply is in the form of a deterministic vector, while the healthcare demand is in the form of a vector of random variables.

- In a medical facility that practices PCMH model, all primary care is performed by numbers of PCMH teams. Designing patient panels and allocating patient population to these multiple teams is a challenge, since the professional mix and staffing level of these teams must balance well with the total workload generated by the entire patient population of the medical facility.

- In any medical facility, due to migration or death, some existing patients drop out from the patient set and some new patients add to the patient population. This necessitates that the patient panels be dynamically updated, and so too the PCMH team staffing levels (which is also susceptible to the similar migration forces).

The goal of this research is to develop a rigorous statistical based workload estimation model by extending the hierarchical generalized linear model to include multivariate responses. This model will provide a good estimate of workload demand portfolio for a relevant set of healthcare professionals for any particular patient based on his/her key demographic, diagnostic and health attributes. In addition we used our proposal on real data from Veteran Healthcare Administration to produce findings that have key public and medical implications.

The remainder of this paper is laid out as follows. Section 2 introduces our data sources and study variables. Section 3 describes the main methodology. Some discussion points and future research directions are presented in Section 4.

II. DATA SOURCE AND STUDY VARIABLES

In this study we collected outpatient data from a random sample of 888 different facilities (which corresponds to 130 VAMCs of all 23 VISNs) during FY11 quarter 3 to FY12 quarter 2. To achieve a better picture of the data environment, we tentatively arranged all independent attributes into five groups as summarized in Table 1[14,15]. It should be noted that these variables remain the same for a patient during the fiscal year.

| *Group* | *Attribute* |
|---|---|
| *Demographic* | **Gender** <br> Male <br> Female |
| | **Age** (as of 7/1/2011, years) |
| | **Marital status** <br> Married <br> Previously married <br> Never married <br> Unknown |
| *Socioeconomic* | **Insurance** (of any types) <br> Yes <br> No |
| | **Employment status** <br> Active Military Service <br> Employed Full-Time <br> Employed Part-Time <br> Not Employed <br> Retired <br> Self Employed <br> Unknown |
| *Enrollment* | **Priority** <br> 1 (service connected disability > 50%) <br> 2 (service connected disability 30%–40%) <br> 3 (service connected disability 20–30%) <br> 4 (catastrophically disabled) <br> 5 (low income or Medicaid) <br> 6 (Agent Orange or Gulf War illness) <br> 7 (non-service connected, income below HUD) <br> 8 (non-service connected, income above HUD) |
| *Utilization* | **VISN** <br> 1 (New England Health Care System) <br> 2 (Network Upstate New York) <br> …. |
| | **Facility** <br> 662 (San Francisco) <br> 537 (Chicago) <br> …. |
| | **PCMH team** |
| | **Assigned provider position** |
| | **Assigned provider experience** (years) |
| | **Changed provider count** |
| | **Provider full time equivalent** |
| | **Length of stay** (inpatient-day) |
| *Clinical* | **Clinical Assessment Need Score** |
| | **Aggregated Condition Category** <br> 1 (infectious and parasitic) <br> 2 (malignant neoplasm) <br> …. |

Table 1: Baseline characteristics of patient factors

The two dependent variables are total primary care and non-primary care Relative Value Units (or RVUs), and for each unique SSN, they are calculated by converting the primary care and non-primary care Current Procedural Terminology (or CPT) codes from all patient visits during

The fiscal year (according to Centers for Medicare and Medicaid Services model). Simply, the Non-PCRVU refers to all of the non-primary care workload during the year, which could be from one or many visits to outpatient specialty care, and the PCRVU is the primary care workload during the year from outpatient primary care.

### III. METHODOLOGY

#### A. Model Specification

The PCMH data is hierarchically organized into three nested levels as shown in Fig.1, where patients are grouped within PCMH teams, and teams are in turn nested within VA facilities. Note that PCMH teams are tied to facilities, i.e., a specific team cannot work at different facilities (teams are nested within facilities). Risk factors can be associated with the response variables at each level while patients from the same team (facility) may have more similar outcomes than patients chosen at random from different teams (facilities). For example, we can study the effects of age (patient-level), PCMH assigned provider's experience (team-level), and type of hospital (facility-level) on the outcomes with nested sources of variability.

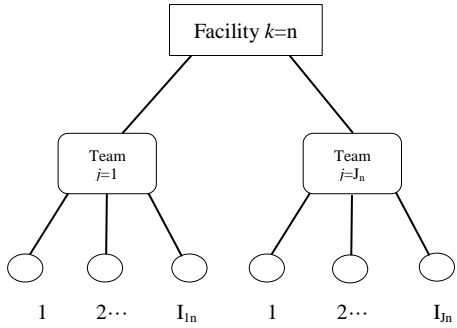

Fig.1: Data structure for PCMH hierarchical model

#### B. Multi-Level Multivariate Analysis

Now a multivariate generalization of this hierarchical GLM is proposed in which both PC and Non-PC workloads are predicted simultaneously. There are several advantages of using a multivariate approach instead of univariate method, Tabachinck et. al. [16]. One is that the multivariate analysis can better control the type I error rate compared to carrying out a series of univariate statistical tests. Second, this approach can shrink the prediction interval of the dependent variables to a large extent when compared to predicting one of them in isolation. Also using a multivariate scheme, the covariance structure of the responses can be decomposed over the separate levels of hierarchy, which can be of much value for multilevel factor analysis [17,18].

Suppose we have P response variables and let $Y_{hijk}$ be the workload on outcome h (PC or Non-PC workload here) of patient i in PCMH team j and facility k. Here we put the measures (responses) on the lowest level of hierarchy, and represent the different outcome variables by defining P dummy variables like

$$d_{phijk} = \begin{cases} 1 & p = h \\ 0 & p \neq h \end{cases} \quad (1)$$

Then we formulate the lowest level as

$$Y_{hijk} = \rho_{1ijk} d_{11ijk} + \rho_{2ijk} d_{22ijk} + \ldots + \rho_{pijk} d_{ppijk} \quad (2)$$

in which neither the usual intercept nor the error term exists as before. The reason for this is that we solely serve the lowest level as a way to define the multivariate structure using dummy variables. Then we may use $\pi$ terms to employ regression equations at the patient level

$$\rho_{pijk} = b_{p0jk} + b_{p1jk} X_{pijk} + e_{pijk} \quad (3)$$

In which a separate index is utilized for denoting the dependent variable of interest. It is noted that with this approach one can fit different intercepts and slopes for different response variables and allow them to vary across any levels of hierarchy. Following (3), at the team level, we can have

$$b_{p0jk} = g_{p00k} + g_{p01k} Z_{jk} + u_{p0jk}$$
$$b_{p1jk} = g_{p10k} + g_{p11k} Z_{jk} + u_{p1jk} \quad (4)$$

Where we introduce our 2-level predictors (level-1 moderators) along with random intercepts and slopes and finally link them to the facility level equations by

$$g_{p00k} = l_{p000} + l_{p001} W_k + u_{p00k}$$
$$g_{p01k} = l_{p010} + l_{p011} W_k + u_{p01k}$$
$$g_{p10k} = l_{p100} + l_{p101} W_k + u_{p10k}$$
$$g_{p11k} = l_{p110} + l_{p111} W_k + u_{p11k} \quad (5)$$

Keeping on this way, one can straightforwardly extend the model to include more predictors at each level and study the effects of fixed and random parameters at any given point. Another advantage of such modeling is that we can impose an equality constraint across all response variables to build a specific relation with certain effects. For example, we can force level-1 regression coefficients for p=1 (PC workload) and p=2 (Non-PC workload) to be equal by adding the constraint $\beta_{11jk} = \beta_{21jk}$. This makes the new model nested within the original model, and thus we can test whether simplifying the model is justified, using a chi-square test on deviances. Plus, if the predictor has random components attached to it, a similar approach would apply to the random part of the model.

IV. ANALYTICS

*A. Model Fitting*

The improvement in model fit is evaluated by DIC over all iterations after the burn-in phase of MCMC simulations. Based on a rule of thumb, we favor the model with lower DIC when the DIC reduction of more than 10 units is observed. Depending on the goodness-of-fit and significance tests, sometimes-intermediate models, such as a reduced version of model 3 with only one significant random slope, are also examined. Performing this strategy, we seek to answer the following three research questions:

- How much of the variance in PC and Non-PC workload is associated with patients, PCMH teams, and VA facilities?
- Does the effect of any patient-level predictor change among PCMH teams or VA facilities? And does the effect of any team-level predictor vary among VA facilities?
- What is the impact of patient non-adherence (as measured by "Changed provider count") on PC workload, controlling for patient, PCMH team, and VA facility characteristics?

Setting the significance level at 0.05, we run the models with 50,000 iterations, a burn-in period of 10,000, and a thinning interval of 25. Although different modeling strategies could be selected for estimating our multilevel model, we focus on the most parsimonious and best-fitting approach for the given data and our specific research questions. To this end, six models (Table 2) from basic to comprehensive are run sequentially and the outputs are reported for each step in order to provide insights for a particular objective. All analyses and computations are done in R version 3.0.2 [*19*]. In order to address the first question, we fit the unconditional model as summarized in Tables 3-5. Note that the first (third) row in each table shows PC (Non PC) intercept variance along with its 95% Highest Posterior Density interval, and the second row corresponds to the workload correlations. The team ICC for the PC outcome is computed as (0.168/(0.609+0.168+0.218)). We find that about 17% of the variation in PC workload exists between PCMH teams and 22% is there between VA facilities, leaving near 61% of the variance to be accounted for by patients. Thus a practically meaningful proportion of all variation happens at higher levels, providing support for our use of a 3-level hierarchical model. These percentages are 5%, 16%, and 79% for Non-PC workload respectively. Interpreting the correlations between PC and Non-PC at different levels can make other useful points. First, the results of a joint conditional independence test Gueorguieva [*20*] show that the RVUs (at the patient level) are positively associated which confirms the fact that a simultaneous modeling of both primary and non-primary care is more reasonable than using one of them in isolation. Second, we infer that the correlation is not significant when it comes to the team level, and it is poorly significant at the facility level. By doing so, we save two DF, but the changes in variance estimates are too trivial to restate here.

*B. Numerical Comparisons*

| Model 1 | Model 2 | Model 3 | Model 4 | Model 5 | Model 6 |
|---|---|---|---|---|---|
| No predictors, just residual and random intercepts (Unconditional) | Model 1 + patient-level predictors | Model 2 + random slopes for patient-level predictors | Model 3 + team-level predictors | Model 4 + random slopes for team-level predictors | Model 5 + facility-level predictors |
| Results used to compute Interclass Correlation Coefficient (ICC) which assesses the degree of clustering among subsets of cases in the data. | Results show the relationships between patient-level predictors and outcomes | Model 2 results + findings that show if the associations between patient-level predictors and the outcomes vary across team-level and facility- level units | Model 3 results + results that reveal the relationships between team-level predictors and the outcomes | Model 4 results + findings that shows if the associations between team-level predictors and the outcomes vary across facility-level units | Model 5 results + results that indicate the relationships between team-level predictors and the outcomes. |

Table 2: Regression modeling strategy and specific results for 3-level hierarchical model

In this section we design three comparison studies to demonstrate some novel aspects of our proposal. First, we evaluate an alternative variance structure. Particularly, for patient (residual), team, and facility random intercepts, we change the parametric matrix to have the same diagonal elements with zero off-diagonals then compare the results. We run each model twice to take control of the Monte Carlo error and keep all other factors constant among different fittings. As shown in Table 3, the best fit is corresponding to the first row in which the proposed variance structure is applied at all levels of hierarchy.

| Facility | Team | Patient | DIC |
|---|---|---|---|
| 2 | 2 | 2 | 225337.8 – 227448.1 |
| 2 | 2 | 1 | 225491.7 – 225494.1 |
| 2 | 1 | 2 | 225401.1 – 225396.9 |
| 2 | 1 | 1 | 225582.5 – 225580.3 |
| 1 | 2 | 2 | 225378.5 – 225375.7 |
| 1 | 2 | 1 | 225444.9 – 225441.2 |
| 1 | 1 | 2 | 225457.8 – 225460.5 |

Table 3: Goodness-of-fit values for the two scenarios

## V. CONLUSION

A key factor in the success of medical homes in delivering quality and coordinated care lies in their teams' ability to handle uncertainties that can be caused by different sources such as patient/physician appointment scheduling, care logistics, and more importantly patients' health demands. This paper addresses the problem of clinical demand prediction in the presence of nested sources of variation at different operational levels. We collected outpatient visit data from a large sample of Veterans Affairs hospitals and investigated the relationship between risk factors at three operational levels and total care demands on a yearly basis. We propose a multivariate multilevel generalized linear model in a Bayesian framework to predict the care demand portfolio in medical home practices. The proposal can fit heteroscedastic variances and unstructured covariance matrices for nested random effects and residuals as well as their interactions with categorical and continuous covariates simultaneously.

Our work can further be extended in some fronts. One challenging direction would be to modify the proposed approach to handle longitudinal observations from past history of care demands for a specific patient profile [21,22]. This may be done by expanding the multivariate distribution of outcomes to include a temporal dimension, which requires great care in model specification and implementations thanks to various inter-correlations [23,24]. Alternatively, one can combine some autoregressive terms to the variance structure introduced in this work. Another issue worth exploring is related to the way that one can adjust for patient risk or comorbidities. Although several algorithms such as Clinical Risk Group (CRG), *veriskhealth* DxCG, and CMS's HCC software have been used in the literature, no scientific study is available to systematically evaluate the impacts of each algorithm on prediction modeling of care demands.